\documentclass[fleqn,12pt,twoside]{article}
\usepackage{espcrc1}

\usepackage{graphicx}
\usepackage[figuresright]{rotating}

\newcommand{\AmS}{{\protect\the\textfont2
  A\kern-.1667em\lower.5ex\hbox{M}\kern-.125emS}}

\title{How the nuclear Fermi motion plus a simple statistical
model explains the EMC effect}

\author{J. Ro\.zynek and G. Wilk,\address{Soltan Institute for Nuclear
Studies, Ho\.za 69, 00-681, Warsaw, Poland}}

\begin{document}

% typeset front matter
\maketitle

\begin{abstract}
We present calculation of influence caused by nucleon Fermi motion
on the parton distributions in nuclei. Our approach is based on
the model where momenta of valence partons have some primordial
distribution inside the hadron at rest, which is either provided
by a statistical considerations or calculated using spherically
symmetric Gaussian distribution with a width derived from the
Heisenberg uncertainty relation. The sea parton contribution emerges 
from the similar Gaussian distribution with a width dictated by the
presence of virtual pions in hadron. We show that the influence of
Fermi motion changes substantially the nucleonic structure function
inside the nucleus in the right direction and therefore should be
considered seriously in all attempts devoted to explain the
experimentally observed EMC effect for $x_{Bj} > 0.1$. 
\end{abstract}

\section{Introduction}
Let us consider deep inelastic scattering of electrons off nuclei
as a following two step process (see Fig. 1): electrons interact with
valence partons (quarks) of nucleons, which in turn interact between
themselves as nuclear constituents and this fact means that their
properties must be suitable modified by nuclear the environment. The
gluonic and the sea $(q\!\!-\!\!\bar{q})$ content of nucleons will be
accounted for by their higher Fock space components \cite{Brodsky} 
of the nuclear wave function like, for example, pions \cite{EI}. The
nuclear structure function is therefore given by the convolution,
\begin{equation}
 F^{A}_{2}(x_{A})/x_{A} = A \int \int dy_{A} dx \delta
(x_{A}-y_{A}x) \rho ^{A}(y_{A}) F^{N}_{2}(x)/x,
\end{equation}
of nucleon distribution function, $\rho^{A}(y_A)$, and structure
function (SF) of free nucleon, $F^{N}_{2}(x)$. Here $x_{A}/A$ (the
Bj\"orken variable for the nucleus) is the ratio of quark and nucleus
longitudinal momenta, $ y_{A}/A$ denotes similar ratio of nucleon
and nuclear longitudinal momenta and $x=x_{Bj}=(E-p_z)/M)$ (the
Bj\"orken variable for the nucleon) is ratio of the quark and nucleon
longitudinal momenta  (with $E=\sqrt{m^2+{p}^2}$ being the parton
energy, $p_z$ its $z$ component and $m$ the nucleon mass).

\section{Results}

We shall demonstrate in the framework of simple statistical model
\cite{Zou} and by using some specific Monte Carlo simulation
\cite{RWA} how SF of free nucleons, $F_2^{(N)}$, changes in nucleus
because of the nuclear Fermi motion (energy momentum conservation is
always strictly imposed). Let us, however, first start with a short 
reminder of the situation encountered in the Nuclear Relativistic
Mean Field (RMF) description of nuclear dynamics. In this approach
electrons are scattered on nucleons moving in constant average scalar
and vector potentials in the rest frame of the nucleus and are
described by Dirac equation,
\begin{equation}
[{\hbox{$\boldmath{\alpha}$}} \cdot {\bf p} + \beta (m+U_{S}) -
(e_{N}-U_{V})]\psi = 0, \label{eq:Dirac}
\end{equation}
with $U_{S}= -g^{2}_{s}/m^{2}_{s} \rho _{s}, U_{V}=V_{\mu }\delta
_{\mu 0}= g^{2}_{v}/m^{2}_{v} \rho $, where $g_i,\ m_i\ (i=s,v)$
are scalar and vector meson coupling constants and their masses,
respectively, whereas $\rho_{s}= \sum^{}_{i}\psi ^{+}_{i}\beta \psi
_{i}$ and $\rho = \sum^{}_{i}\psi ^{+}_{i}\psi _{i}$ denote the
respective scalar and (fourth component) vector densities. The
positive and negative solutions of eq. (\ref{eq:Dirac}) are used.
It turns out that to explain the observed enhancement of the
spin-orbit part of nucleon-nucleus optical potential one needs
relatively big values of $U_S=300$ MeV and $U_V=400$ MeV. On the
other hand it can be shown \cite{MEANFIELD} that nucleon distribution
function inside the nucleus can be written as:
\begin{equation}
\rho ^{A}(y_{A}) = {3\over 4} (v^{2}_{A}-(y_{A}-1)^{2})/v^{3}_{A},
\end{equation}
where  $v_{A}=p_{F}/E_{F}$ and $y$ is limited to $0<(E_{F}^{^{*}}
-p_{F})<my<(E_{F}+p_{F})$. $E_{F}$ is the nucleon Fermi energy and
the nucleon chemical potential $\mu = m - 8$ MeV. The corresponding
results (dashed line in Fig. 2) show that in such approach the
influence of nuclear medium disappears \cite{MEANFIELD} (one 
should also mention at this point that the widely discussed
contributions from the additional, i.e., in the medium, pions were
recently disclaimed \cite{NUCLPION}). The reason is that in our
calculations leading to this result the quark primordial distribution
(PD) of the nucleon in the medium is left unchanged. In what follows
we shall now examine how the nuclear medium can change this PD using
two different phenomenological approaches.

\subsection{Simple Statistical Model}

We start with statistical model for nucleon structure functions in
which partons are considered as a gas of noninteracting particles
remaining on energy shell \cite{Zou} (i.e., with 
$E_{i}\equiv\sqrt{\overrightarrow{p^{2}}+m_{i}^{2}} =
|\overrightarrow{p^{2}}|$) with the following primordial momentum
distribution:
\begin{equation}
d\varrho(p_{1}..p_{n}) \sim
\delta^{4}(P-\sum_{i}p_{i})\prod_{i}dE_{i}d\Omega_{i}
\end{equation}
(because of the confinement the free-parton distribution was divided
here by $E_{i}$ \cite{Zou}). For nucleons in the nuclear medium we
shall separate out the Fermi motion and write the average (over
nucleus) sum of the energies of all partons as
\begin{equation}
\sum_{i}E_{i}/A = \mu = 0.6\cdot \langle E_{F}\rangle +
E^{Relative}_{N}\quad \Longrightarrow 
\quad \langle
 E^{Relative}_{parton}\rangle = E_{i} - 0.6\cdot \langle
E_{F}\rangle/n  \label{eq:Fermi} ,
\end{equation}
where $0.6\cdot \langle E^{Fermi}_{N}\rangle$ is the average kinetic
(Fermi) energy of nucleon and $E^{Relative}_{N}$ is the nucleon
energy in its rest frame. In this way we are introducing a change in
the parton energy (by shifting it to smaller values
(\ref{eq:Fermi})), which, in turn, results in the width of energy PD
being smaller than in the analogous distribution for the free nucleon (and,
at the same time, being more peaked at the origin). This results in
the depletion of $F^{N}_{2}(x)$ in nuclear medium for $x>0.2$.
However, for $x<0.2$ the scale $z>1/Mx$ shown in Fig. 1 becomes
comparable with the actual size of nucleon (or bigger that it, the
expected sizes of corresponding nuclear clusters also grow
accordingly). Therefore in this region the correction
(\ref{eq:Fermi}) deduced only from the single nucleon Fermi motion
in not longer valid \cite{parnotprob} and we have to consider also
motion of the whole interacting cluster with size given by
$A^{1/3}\sim1/x$, which, in the limit of $x=0$, becomes a whole
nucleus . In another words: for $x<0.2$ the time of the
electron-nucleus interaction is sufficiently large for the parton,
which has been hit, to change in the final state (because of the 
collective nucleon-nucleon interactions) both its localization and
its four momentum. The results taking this effect into account are
shown in Fig. 2 as solid line. 

\subsection{Monte Carlo Modelling with Pure Fermi Motion.}

We proceed now to the next phenomenological example in which we are
using the simple Monte Carlo model developed recently \cite{EI}. Let
$j$ denote four-momentum of the struck parton (probed by current with
virtuality $Q^2_0$) selected (for valence quarks) from Gaussian PD
with width $0.172$ GeV, $r$ - the respective four-momentum of
hadronic remnants and $W$ and $W'$ their respective invariant masses.
Only events satisfying the exact kinematical constraints of the
corresponding deep-inelastic reaction probing our nuclear
distribution, 
\begin{equation}
0 \leq j^2 \leq W^2 ,\qquad \qquad  0 \leq r^2 \leq W'^2 \ ,
\end{equation}
are accepted and selected to form the final distribution we are
looking for. Calculating the momentum distribution of partons inside
bound nucleons, with our basic assumption that primordial parton
distribution in the nuclear medium remains the same but that we
include now Fermi motion of nucleons, we have to subtract the Fermi
motion in the nucleon rest frame. This mechanism works similarly to
the decreasing of the size of the valence parton momentum
distribution used in \cite{RWA} and produces both the minimum for $x$
around $x \sim 0.6$ and the maximum around $x \sim 0.1$. It can be
interpreted as increasing de-confinement region in the configuration
space. As in the statistical model above (\ref{eq:Fermi}), the 
substraction of the Fermi average energy changes the energy width of
energy parton PD by $~7\%$. The corresponding results are presented
in Fig. 2 as full triangles (which practically follow the solid line
representing results of the previous approach showing therefore their
equivalence). The sea parton
distribution is given here by the convolution of the pionic component
of the nucleon, $f_\pi(x;Q^2_0)$, and the parton structure of pion,
$f_{pion}(x;Q^2_0)$, obtained from the same Gaussian PD as used for
valence partons. Notice that the influence of Fermi motion is strong
but the agreement for small $x$ is lost. 

\subsection{Full Monte Carlo Modelling}

We shall therefore improve the treatment of sea quarks. In the 
presented model the sea parton distribution is generated directly 
from distribution of the pionic cloud which surrounds the nucleon
core constituted by the valence quarks. Because part of these pions
is responsible for mediating nucleon-nucleon interactions, the
corresponding part of the sea quarks must be connected rather to
whole nuclear medium and not to the individual nucleons. For small $x$
the crucial factor turns out to be the change of the nuclear virtual
pion cloud connected with the exchanged mesons responsible for the
the nuclear forces \cite{RWA}. In order to be able to fit data 
in this region we have therefore to adjust the value of the parameter
which determines the relative number of the (effective) intermediate
pions (which are assumed to mediate the nucleon-nucleon interaction).
It turns out \cite{RWA} that the proportion leading to good results
is to assign $93\%$ of virtual pions to contribute to the sea quark
structure function of the nucleon and let the rest to be responsible
for the the nucleon-nucleon interaction.  Full calculation in which
the width of the energy-momentum PD was changed by $10\%$ (from
$0.18$ GeV to $0.165$ GeV) result in very good fit to experimental
data including the small $x$ region (see Fig. 2 - full squares).

\vspace*{-.9cm} \hspace{-2cm}
\begin{figure}[h]
\begin{minipage}[t]{0.475\linewidth}
\centering
\includegraphics[height=4cm,width=4.4cm]{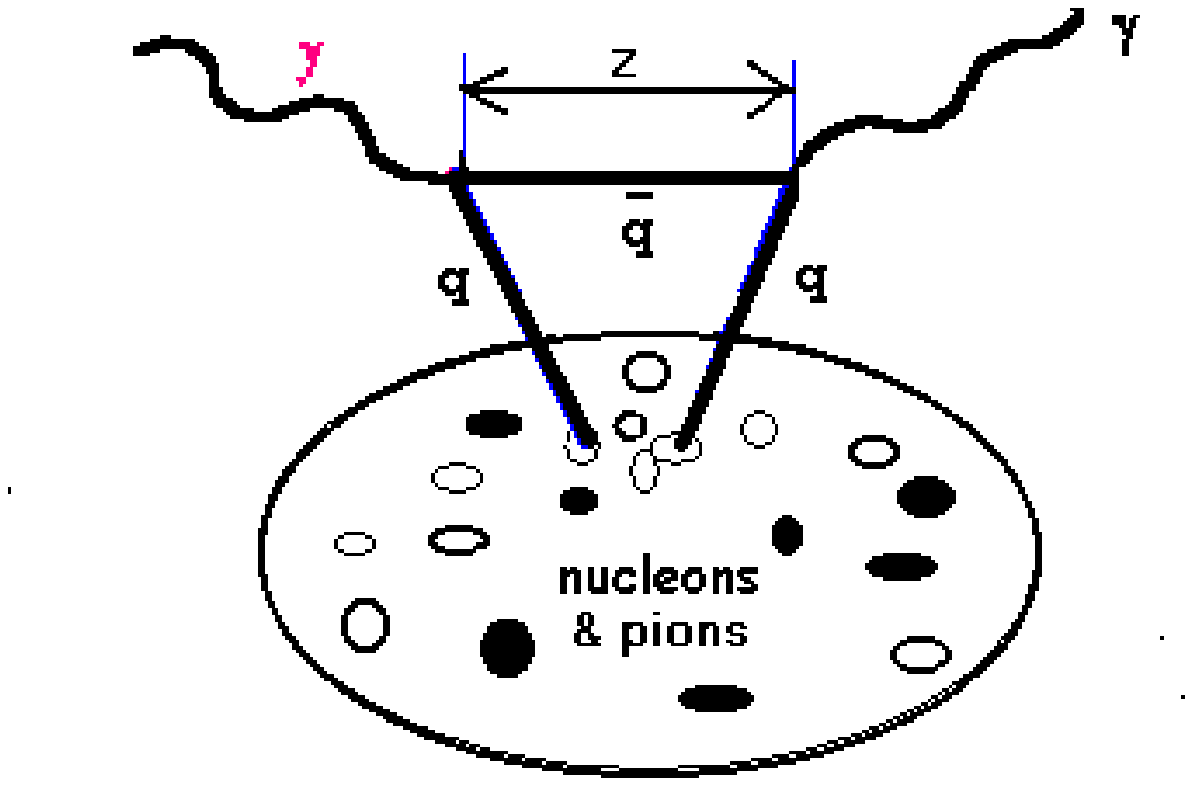}
\caption{Struck quark (from nucleon or
 pion) propagates through the nucleus.}
\end{minipage}\hspace{-1mm}
\begin{minipage}[t]{0.475\linewidth}
\centering \vspace{-53mm} \hspace{15mm}
\includegraphics[height=6.6cm,width=9.5cm]{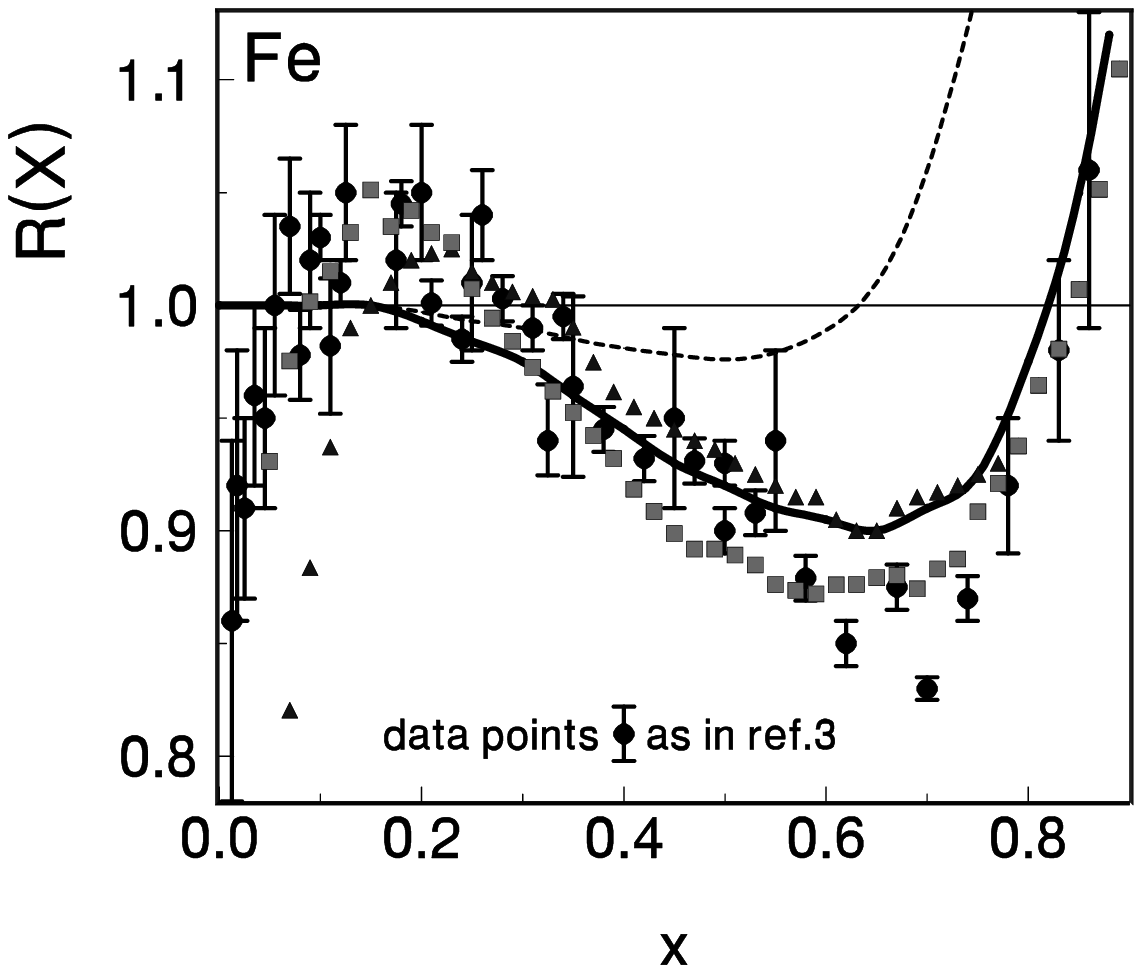}
\caption{R(x) - the ratio of nuclear structure functions (iron to
deuterium).}
\end{minipage}
\end{figure}
\vspace{-1cm}

\section{Conclusions}
We conclude by stating that Fermi motion of nucleons inside the
nucleus provides very big contribution to the nucleon PD in the
nuclear medium (at least in the examples considered here). Therefore,
contrary to some previous calculations and conclusions to the
opposite \cite{MEANFIELD} (including recent work \cite{MILLER}),
it affects strongly the observed EMC ratio in the broad range of
variable $x > 0.1$. \\

\noindent This work has been supported in part by Grant
2-P03B-048-12 of the Committee for Scientific Research.

\end{document}